\documentclass[%
preprint,
 amsmath,amssymb,
 aps,
prb,
showkeys
]{revtex4-2}

\usepackage{graphicx}
\usepackage{dcolumn}
\usepackage{bm}

\begin{document}


\title{Lorentz force on superconducting vortices near line defects}

\author{Ruby A. Shi$^{1}$}
\altaffiliation{Now at Quantum circuits, Inc.}
\email{rubyshi@stanford.edu}
\affiliation{$^1$Department of Physics, Stanford University}

\date{\today}

\begin{abstract}
In type-II superconductors, magnetic flux penetrates in the form of quantized vortices whose dissipative motion, driven by the Lorentz force, can degrade superconductivity. Understanding vortex dynamics in both homogeneous regions and near unavoidable structural defects is crucial for superconducting applications. This study examines a scenario in which a superconducting quantum interference device (SQUID) scans across a thin-film superconductor containing line defects. We first estimate the radial Lorentz force on a vortex in a homogeneous region using both analytical methods and numerical simulations based on the fast Fourier transform algorithm. For a film with a Pearl length of 400 micrometers and a SQUID height of 4 micrometers, we find that the SQUID tip can exert a force of approximately 3 femtonewtons on a vortex. We then evaluate the Lorentz force on vortices near two parallel line defects. Our results show that the Lorentz force is enhanced for vortices pinned on or between line defects. Vortices pinned on the line defects experience force enhancement predominantly perpendicular to the defects, while vortices in between experience enhancement along the defect direction. These findings enable more accurate estimation of Lorentz forces on vortices near line defects in thin-film superconductors and contribute to the broader understanding of vortex pinning and dynamics in defect-engineered superconductors. The methods can be extended to bulk superconductors and generalized to other defect geometries.
\end{abstract}


\maketitle


\section{Motivation}

Superconducting vortices in type-II superconductors are whirlpools of supercurrents circulating around a normal (non-superconducting) core, each carrying a single flux quantum, $\Phi_0$ \cite{tinkham, blatter_review}. Vortices tend to anchor their cores at naturally occurring defects such as atomic vacancies or twin boundaries. When a current flows through the superconductor, it exerts a Lorentz force on the vortices, driving their motion. This motion induces dissipation and ultimately destroys the superconducting state \cite{brandt1995, yeshurun1996}. Consequently, significant effort has been devoted to pinning vortices at engineered defects, known as artificial pinning centers, to suppress vortex motion and enhance the critical temperature $T_c$ \cite{civale1991,matsuura2021}. A detailed understanding of vortex dynamics, in terms of the magnitude and direction of the pinning force, is therefore essential for developing practical superconducting technologies \cite{larbalestier2001, gurevich2011}.
\par
The magnitude and direction of the Lorentz force acting on isolated vortices can be revealed by local magnetic probes, such as scanning superconducting quantum interference devices (SQUIDs) \cite{John_squid}. In superconductors, vortex pinning centers are often modeled as local minima of a pinning potential, where the shape and depth of the potential determine the depinning force—the minimum force required to dislodge a vortex either partially or completely from its pinned position. Magnetic scanning probes can exert a controllable Lorentz force on vortices and image the resulting flux line distortions, offering insight into vortex dynamics \cite{Straver2008Controlled, auslaender2009mechanics, zhang2015single}. In particular, a scanning SQUID equipped with a field coil can apply localized Lorentz forces and simultaneously image the vortex response, enabling quantitative mapping of pinning landscapes and depinning thresholds \cite{Luan2010Local, IreneFeSe}.
\par
The vortex dynamics in thin-film superconductors with zero-dimensional atomic point defects were studied by Noad et al. \cite{HilaryDefect}. Their work first provided a theoretical framework for how atomic-scale defects modify the Lorentz force experienced by pinned vortices. They then experimentally demonstrated that these modifications manifest in scanning SQUID susceptometry measurements as localized "halos." Specifically, the magnetic field from the SQUID field coil applies a Lorentz force to the vortex, distorting its flux lines. This distortion is detected by the SQUID’s pickup loop. The study showed that atomic or mesoscopic point defects, significantly smaller than the pickup loop size, can be effectively modeled as localized suppressions of the superfluid density. This work highlights scanning SQUID susceptometry as a powerful technique for detecting and characterizing local variations in superfluid density. However, the mathematical treatment in \cite{HilaryDefect} employed a simplified model in which the point defect was represented as a radially symmetric delta function. In this manuscript, the modification of the Lorentz force in thin films containing two parallel line defects is calculated. These one-dimensional defects serve as a minimal model for twin boundaries, which are commonly observed in high-temperature superconductors and are of practical interest \cite{beyer1989twinning, jiang1988microstructures, twinNb2023, watashige2012suppression, pina2006twin}. The results show that the Lorentz force modifications are largely confined to regions near the line defects, supporting the validity of the two-line model as a minimal yet representative approximation, despite the fact that real twin boundaries typically consist of multiple, closely spaced parallel line defects.

\section{Lorentz force by SQUID in homogeneous superconductors}
\begin{figure}
    \centering
    \includegraphics[scale = .35]{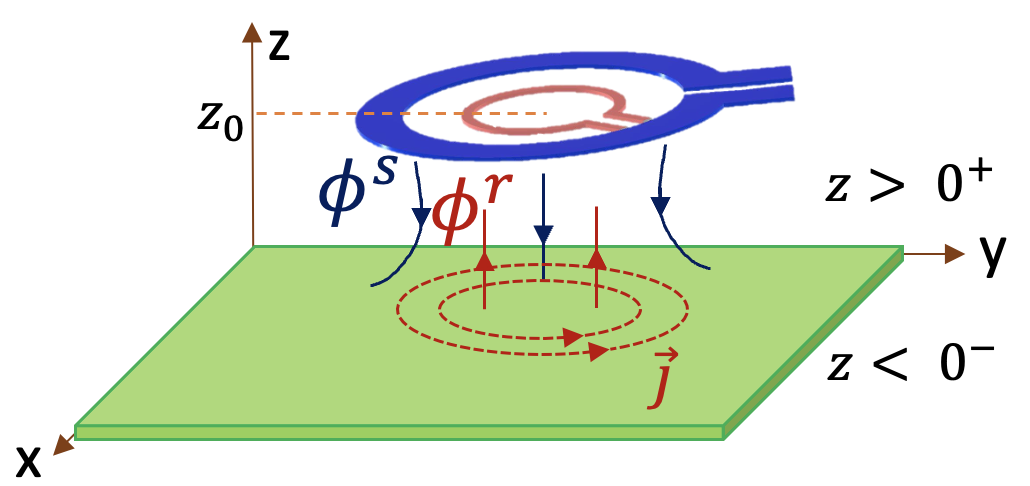}
    \caption{Schematic representation of the problem setup.}
    \label{schematic}
\end{figure}

The Lorentz force exerted on vortices by the field coil in a homogeneous thin-film superconductor, schematically illustrated in Fig.~\ref{schematic}, is first considered. A thin-film superconductor (green sheet) is positioned in the $xy$-plane at $z = 0$, with its top surface at $z = 0^+$ and bottom surface at $z = 0^-$. The concentric pickup and field coil loops (pink-blue loop pair) are located above the film in the positive half-space at a height $z_0$. Magnetic field lines generated by the field coil are illustrated in blue. In regions outside the superconductor, where no bound current exists, this magnetic field can be expressed as the gradient of a source scalar potential $\phi^s$. In response to the applied field, the superconductor generates a screening circular current $\vec{j}$, represented by red dashed lines. This screening current exerts a Lorentz force on a vortex, for example, one pinned at the origin $(0,0,0)$, which in turn creates a response field that opposes the applied field. Above the thin film, where no current is present, the resulting response field and the distortion of the vortex flux line can also be described as the gradient of a scalar potential $\phi^r$.
\par
The Lorentz force on a vortex with flux lines oriented along the $z$-direction is given by
\begin{equation}
    F = \Phi_0 \vec{j} \times \hat{z}
    \label{ForceGeneral}
\end{equation}
where $\vec{j}$ is the supercurrent density, $\hat{z}$ is the unit vector in the $z$-direction, and $\Phi_0$ is the magnetic flux quantum. Kogan showed that in thin-film superconductors, the supercurrent density $\vec{j}$ can be expressed in terms of derivatives of the magnetic field $\vec{h}$ \cite{koganThinFilmDefine}.

By simplifying the field coil and pickup loop pair to two concentric loops, the Lorentz force on a vortex in a thin-film superconductor (in SI units) was derived by Kogan \cite{KoganForce} as:
\begin{equation}
    F(r) = - \Phi_0 I a \int ^{\infty}_{0} dk \frac{ke^{-k z_0}}{1 + \Lambda k} J_1(ka)J_1(kr)
    \label{ForceAnalytical}
\end{equation}
where $\Phi_0$, $I$, and $a$ denote the flux quantum, field coil current, and effective coil radius, respectively. The variable $r$ represents the lateral distance from the center of the field coil to the vortex. The Lorentz force $F(r)$ is radial and expressed as an integral over momentum space. Related estimations for the depinning force in bulk superconductors using similar expressions can be found in \cite{BeenaTwin, ForceYBCO}. It should be noted that $F(r)$ has a sign opposite to $r$, indicating that the force is attractive, or a dragging force, as the SQUID scans across the sample. 
\par
A detailed derivation of (\ref{ForceAnalytical}) can be found in Sections 4.1.1 to 4.1.3 in Shi's doctoral thesis \cite{thesis}. The general idea is to integrate (\ref{ForceGeneral}) over the film thickness and relate the in-plane superfluid density $\vec{j}$ to the shear components of the magnetic field $h$ using the London’s equation
\begin{equation}
    \vec{h} + \lambda^2 \nabla \times \vec{j} = 0
    \label{London}
\end{equation}
where $\lambda$ is the London penetration depth. In the thin film limit, the Pearl length $\Lambda$ is introduced as an effective screening length, given by $\Lambda = 2 \lambda^2/d$, with $d$ being the film thickness. 
\par
The thin film case is simpler than the bulk scenario due to the absence of $z$-dependence in the magnetic field $\vec{h}$. Instead, the tangential components of $\vec{h}$ exhibit a discontinuity across the film: $h_{x,y}(x, y, 0^+) = - h_{x,y}(x, y, 0^-)$. Additionally, the current density $\vec{j}$ can be assumed to be uniform along the $z$-axis, consistent with Pearl’s original treatment of thin film superconductors as two-dimensional current sheets \cite{PearlOrigional}. Due to the radial symmetry of the Lorentz force in the homogeneous case, it is sufficient to calculate only the force component in the $x$-direction:
\begin{equation}
F_x(x, y, 0^+) = \Phi_0 \int^{0^+}_{0^-}dzj_y = \Phi_0 \int^{0^+}_{0^-}dz\frac{\partial}{\partial z}h_x^r = 2 \Phi_0 h^r_x(x,y,0^+)
\label{ForceDirection}
\end{equation}
The response field $h_x^r(x, y, 0^+)$ can be written as a Fourier sum of $h_x^r(\vec{k}, 0^+)$, which can also be written as the gradient of a scalar potential $ik_x \phi^r(\vec{k}, 0^+)$ in the two-dimensional Fourier space.  In the thin film limit, the response scalar potential is related to the source potential by $\phi^r(\vec{k}, 0^+) = \phi^s(\vec{k}, 0^+) /(1 + k\Lambda)$. Thus, the Lorentz force calculation reduces to determining the appropriate $\phi_r$ whose gradient in the  x-direction gives the field $h_x^r$. The procedure for obtaining the Lorentz force in the y-direction is analogous, with the subscript $x$ replaced by $y$.
\par
The Lorentz force can be evaluated using two complementary approaches. The first approach involves directly evaluating the analytical expression provided in Eq.~(\ref{ForceAnalytical}). The second approach entails computing the response field component $h_x^r(\vec{k}, 0^+)$ in Fourier space, followed by a numerical inverse Fourier transform to recover the real-space field $h_x^r(x, y, 0^+)$.

\begin{figure}
    \centering
    \includegraphics[scale = .65]{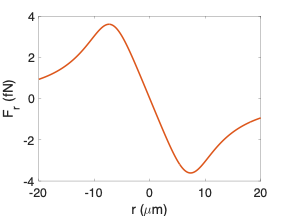}
    \caption{Analytical calculation of the Lorentz force acting on a vortex with Pearl length $\Lambda = 420$, with the SQUID positioned 4.2 $\mu$m above the superconducting film.}
    \label{analytical}
\end{figure}

Figure~\ref{analytical} presents the Lorentz force acting on a vortex with Pearl length $\Lambda =$ 420 $\mu$m, computed using Eq.~\ref{ForceAnalytical} for a SQUID positioned 4.2 $\mu$m above the film surface. The SQUID is centered at $y=0$. The force direction is observed to be opposite to the SQUID–vortex displacement vector $\vec{r}$, indicating that the interaction is attractive.

\begin{figure}
    \centering
    \includegraphics[scale = .45]{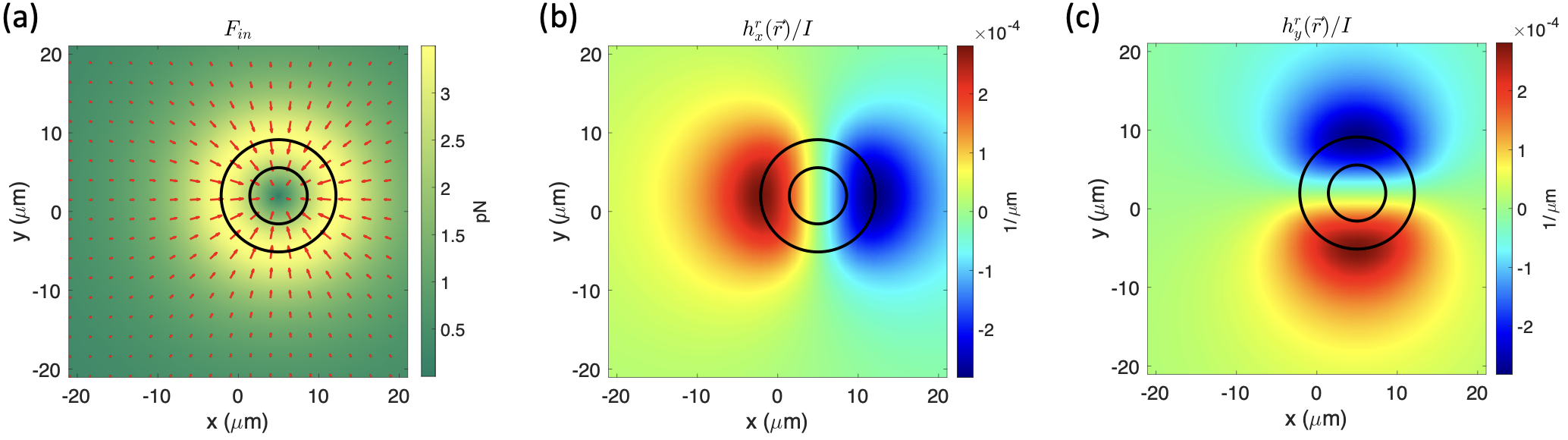}
    \caption{Calculation of the Lorentz force using the inverse Fourier transform method. The black circles denote the locations of the pickup loop and field coil. The SQUID is positioned 4.2 $\mu$m above a superconducting film with Pearl length $\Lambda$ = 420 $\mu$m. (a) Spatial map of the Lorentz force acting on a vortex. The color scale represents the force magnitude, while red arrows indicate the force direction. The xy-coordinates correspond to the vortex position. As expected, the Lorentz force exhibits radial symmetry about the center of the field coil–pickup loop pair. (b, c) Normalized magnetic field component $h_x/I$, $h_y/I$. These field components determine the Lorentz force in the corresponding directions via Eq.~(\ref{ForceDirection}).}
    \label{FFTMethod}
\end{figure}

Alternatively, the Lorentz force can be computed analytically using MATLAB’s inverse fast Fourier transform (FFT) algorithm. Figure~\ref{FFTMethod}(a) displays the magnitude of the Lorentz force, with arrows indicating its direction. The SQUID is not centered at
(0,0), but rather at the center of the black loops denoting the pickup and field coil pair. The parameters used in this calculation ($\Lambda$ = 420 $\mu$m, $z_0$ = 4.3 $\mu$m) are identical to those in Fig.~\ref{analytical}. Figure~\ref{FFTMethod}(b,c) present the magnetic field components $h_x^r$ and $h_y^r$, respectively. These components are directly related to the Lorentz force via equation (\ref{ForceDirection}). 
\par
The results obtained from the two methods show excellent agreement, as expected. However, two important considerations are necessary to ensure accurate computation: 1. Particular care must be taken in defining the k-space, especially at the origin ($k=0$). It is advisable to verify the correctness of the grid by applying the Fourier transform procedure to a function with a known analytical transform. 2. The extent of the real-space grid must be sufficiently large—ideally greater than the Pearl length to properly resolve the singularity at $k = -1/\Lambda$ in the response potential $\phi^r = \phi^s/(1 + k\Lambda)$. More details can be found in Ruby Shi's thesis \cite{thesis}. 

\section{Lorentz force by SQUID in superconductors with line defects}
Diamagnetic screening inhomogeneity is a common feature in superconductors. Some forms of inhomogeneity can be mitigated through improved sample synthesis, while others are intrinsic to the material due to structural defects. It has been shown that the superfluid density may be suppressed along twin boundaries in cuprates and enhanced in pnictides, thereby influencing vortex pinning and dynamics\cite{BeenaTwin,LoganFridge}. In previous literature, twin boundaries have been modeled as lines with locally modified Pearl lengths\cite{KoganLine}.
\par
This section investigates the Lorentz force on a vortex situated between two such lines of modified Pearl length. The results may provide insight into anisotropic Lorentz forces experienced by vortices confined on and between twin boundaries. 
\par
In reference \cite{KoganLine}, Kogan and Kirtley modeled a line-shaped superfluid density defect located at $x = x'$ as a delta-function perturbation, enabling analytical treatment of its impact on vortex behavior.
\begin{equation}
    \Lambda(x) =\Lambda_0 - \alpha^2 \delta(x - x')
\end{equation}

Here, the constant $\alpha$, with units of length, characterizes the strength of the defect. For a spatially inhomogeneous Pearl length $\Lambda(x,y)$, the London equation in the z-direction is modified accordingly to account for the variation in superfluid screening \cite{HilaryDefect}.

\begin{equation}
    h_z^s + h_z^r + \Lambda \left(\frac{\partial h_x^r}{\partial x} + \frac{\partial h_y^r}{\partial y} \right) + \left( h_x^r\frac{\partial \Lambda}{\partial x} + h_y^r\frac{\partial \Lambda}{\partial y}\right) = 0
\end{equation}
\par
It can be derived that the response potential takes the form 
\begin{equation}
    \phi^r = \frac{\phi^s}{1 + k \Lambda}+ \delta \phi^r
\end{equation}
\par
The first term corresponds to the response potential in the isotropic case, while the second term represents the correction arising from the presence of the line defect.
\begin{equation}
    \delta \phi^r (\vec{k}, x_0, y_0) = \frac{\alpha^2}{k(1 + k\Lambda_0)}\int^{\infty}_{\infty} \frac{dq_x}{2\pi} \frac{(\vec{k} \cdot \vec{Q})\phi^s(\vec{Q}, x_0, y_0)e^{-i(k_x - q_x)x'}}{1 + \Lambda_0 Q}
    \label{DeltaOneLine}
\end{equation}
\par
Here, $\vec{Q} = (q_x, k_y)$ is the wave vector in the Fourier space, and $\phi^s(\vec{Q}, x_0, y_0)$ represents the source potential at ($x_0$, $y_0$). Kogan derived equation (\ref{DeltaOneLine}) for $y_0 = x' = 0$. Ruby Shi's thesis shows derivation of equation (\ref{DeltaOneLine}) for general coordinates. 
\par
The source potential on the superconducting film due to a field coil carrying current $I$, with effective radius $a$, positioned at $\vec{r_0} = (x_0, y_0)$ and height $z_0$ above the film, is given by: 
\begin{equation}
\phi^s(\vec{k}, 0) = \frac{\pi I a}{k} e^{-kz_0 -i\vec{k} \cdot \vec{r_0}}J_1(ka)
    \label{sourcePotential}
\end{equation}
\nocite{*}
\par
Building on Kogan’s previous derivation, this analysis focuses on the Lorentz force acting on a vortex situated between two line defects. The system consists of a superconducting thin film lying in the xy-plane at $z = 0$. Two line defects are placed at positions $x = x'$ and $x = -x'$, and the Lorentz force on a vortex within the simulation region is evaluated. This force results from the screening currents induced by a field coil located at $(x_0, y_0, z_0)$. The presence of line defects alters the superfluid response, introducing anisotropy into the screening currents and consequently into the Lorentz force.
\begin{equation}
    \Lambda(x) =\Lambda_0 - \alpha^2 \delta(x - x') - \alpha^2 \delta(x + x')
\end{equation}
\par
The line defects are assumed to have identical strength $\alpha$, consistent with typical modeling of twin boundaries\cite{KoganLine}. Under this assumption, the additional term in the response function is modified accordingly to:
\begin{equation}
    \delta \phi^r (\vec{k}, x_0, y_0) = \frac{2\alpha^2}{k(1 + k\Lambda_0)}\int^{\infty}_{\infty} \frac{dq_x}{2\pi} \frac{(\vec{k} \cdot \vec{Q})\phi^s(\vec{Q}, x_0, y_0)cos((k_x - q_x)x')}{1 + \Lambda_0 Q}
\end{equation}
substitute $\phi^s$ by (\ref{sourcePotential})
\begin{equation}
    \delta \phi^r (\vec{k}, x_0, y_0) = \frac{\alpha^2 I a}{k(1 + k\Lambda_0)}\int^{\infty}_{\infty} dq_x \frac{(\vec{k} \cdot \vec{Q})e^{-Qz_0 - i q_x x_0 - i k_y y_0}J_1(Qa)cos((k_x - q_x)x')}{(1 + \Lambda_0 Q)Q}
    \label{DeltaTwoLines}
\end{equation}
Due to translational symmetry along the $y$-axis, it is sufficient to study $F_x$ and $F_y$ only along a line cut $y_0 = 0$. First, we study $\delta F_x$ at $y_0 = 0$. The additional term to field $h_x$ due to the line defects is 
\begin{equation}
    \delta h^r_x (\vec{k}, x_0, 0) = \frac{i k_x \alpha^2 I a}{k(1 + k\Lambda_0)}\int^{\infty}_{\infty} dq_x \frac{(\vec{k} \cdot \vec{Q})e^{-Qz_0 - i q_x x_0}J_1(Qa)cos((k_x - q_x)x')}{(1 + \Lambda_0 Q)Q}
    \label{ForceTwoLinesX}
\end{equation}
\begin{figure}
    \centering
    \includegraphics[scale = .55]{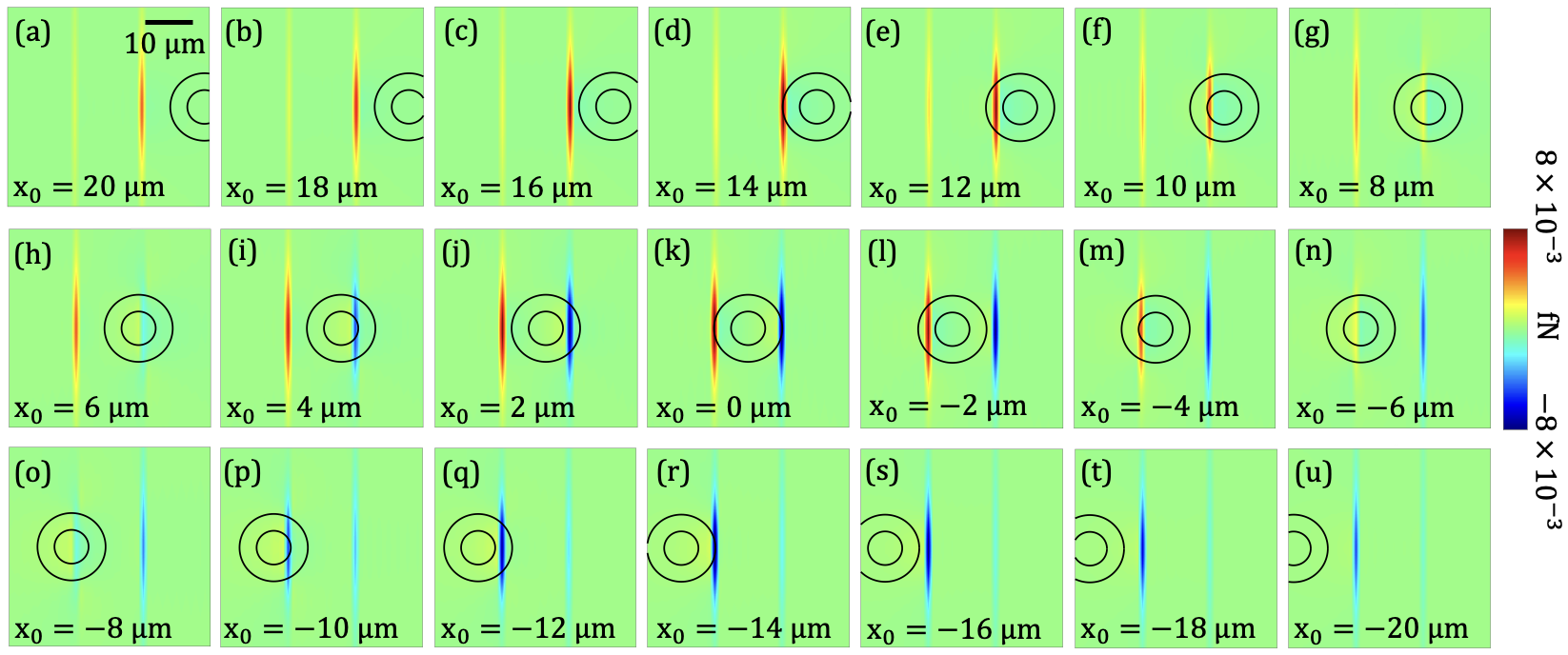}
    \caption{Simulated additional Lorentz force $\delta F_x(\vec{r})$ in the x-direction on a vortex in the presence of two line-shaped superfluid density defects. Color intensity at each pixel represents the magnitude of the x-component of the Lorentz force additional to what's shown in Fig.\ref{FFTMethod}(b) acting on a vortex pinned at that location. The simulation parameters are: defect strength $\alpha = 1~\mu\text{m}$, defect positions at $x = \pm 7~\mu\text{m}$, field coil current $I = 3~\text{mA}$, Pearl length $\Lambda_0 = 400~\mu\text{m}$, and coil height $z_0 = 4~\mu\text{m}$ above the superconducting film.}
    \label{xLoop}
\end{figure}
The additional field component $\delta h_x^r(\vec{k})$ in momentum space depends on $x_0$, the lateral position of the field coil. By numerically applying an inverse Fourier transform to $\delta h_x^r(\vec{k})$, one obtains the corresponding real-space field $\delta h_x^r(\vec{r})$, from which the x-component of the Lorentz force can be directly computed using Eq.(\ref{ForceDirection}) for a given $x_0$. Figure \ref{xLoop} shows simulation results at various coil positions, where the color depth of each pixel indicates the x-component of the Lorentz force acting on a vortex pinned at that location. Two observations follow. First, for vortices pinned on line-shaped superfluid density enhancements (modeled as a positive $\alpha$), the SQUID applies a stronger attractive Lorentz force on the vortex towards the field coil center, consistent with the attractive Lorentz force seen in Fig.~\ref{FFTMethod}(b). This indicates that vortices pinned on pin boundaries are dragged more strongly by the SQUID in the direction perpendicular to the twin boundaries. This result warrants caution when the depinning force is inferred from SQUID dragging experiments \cite{BeenaTwin} and establishes SQUIDs as promising tools to study twin boundary strength. Second, for vortices not pinned on line defects, the modification to the Lorentz force is negligible outside the defect regions, particularly in our case where the two line defects are separated by 14 $\mu$m. Thus, estimating the depinning force using an isotropic model remains valid for vortices located away from the defect sites.
\par
Second, we study $F_y$ along $y_0$ = 0 obtained similarly from the field $\delta h^r_y$,
\begin{equation}
    \delta h^r_y (\vec{k}, x_0, 0) = \frac{i k_y \alpha^2 I a}{k(1 + k\Lambda_0)}\int^{\infty}_{\infty} dq_x \frac{(\vec{k} \cdot \vec{Q})e^{-Qz_0 - i q_x x_0}J_1(Qa)cos((k_x - q_x)x')}{(1 + \Lambda_0 Q)Q}
    \label{ForceTwoLinesY}
\end{equation}
\par
\begin{figure}
    \centering
    \includegraphics[scale = .38]{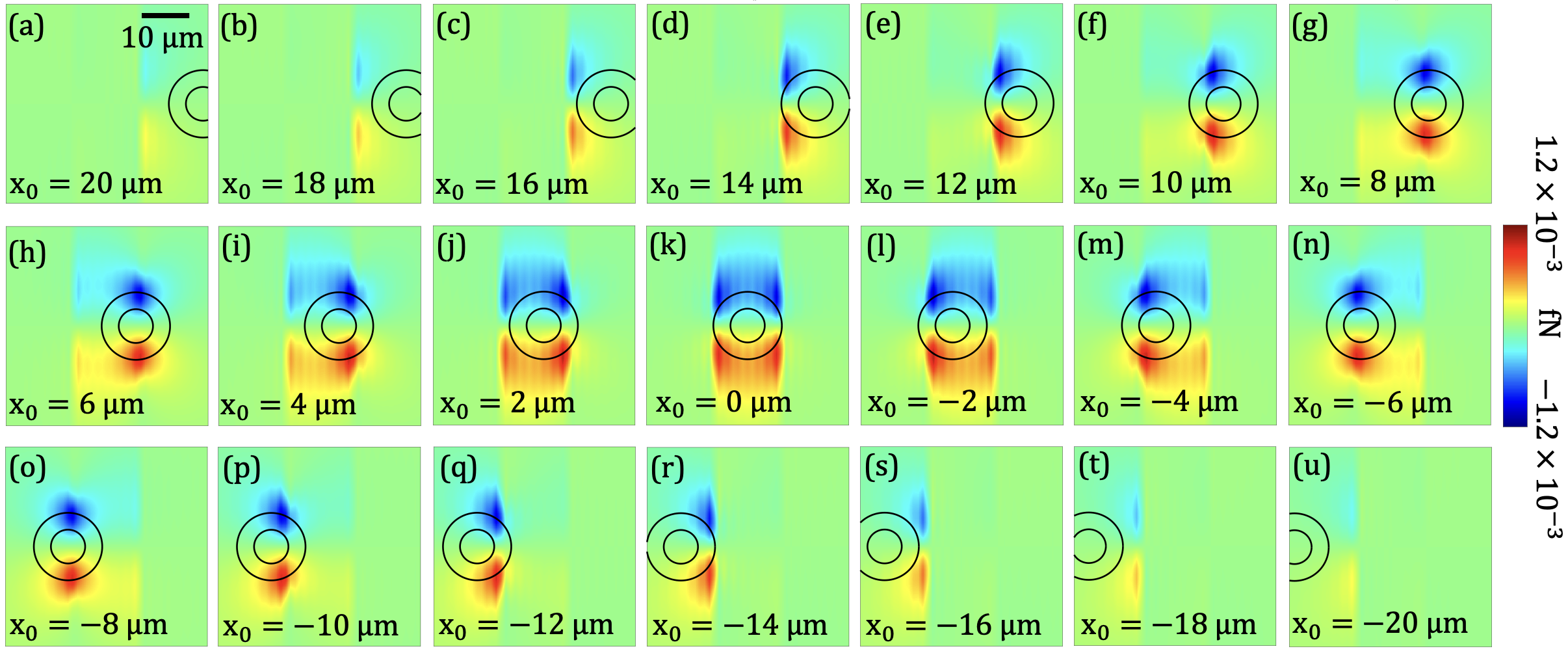}
    \caption{Simulated additional Lorentz force in the y-direction, $\delta F_y(\vec{r})$, on a vortex in the presence of two line-shaped superfluid density defects. The color intensity at each pixel represents the magnitude of the y-component of the Lorentz force acting on a vortex pinned at that location. Simulation parameters: defect strength $\alpha = 1~\mu\text{m}$, field coil current $I = 3~\text{mA}$, defect positions at $x = \pm 7~\mu\text{m}$, Pearl length $\Lambda_0 = 400~\mu\text{m}$, and coil height $z_0 = 4~\mu\text{m}$ above the superconducting film.}
    \label{YLoop}
\end{figure}
Figure~\ref{YLoop} shows the simulated additional Lorentz force in the y-direction on a vortex as the SQUID scans horizontally across the sample. The color depth at each pixel represents the magnitude of the y-component of the force experienced by a vortex pinned at that location, while the black circles indicate the SQUID position. The striped patterns visible in Fig.~\ref{YLoop}(i)--(m) are numerical artifacts resulting from coarse spatial resolution used to reduce simulation time. The results show that vortices pinned between the line defects experience an enhanced y-directional dragging force, with the maximum occurring when the SQUID is directly above or between the defects. This behavior resembles the x-directional force enhancement but is approximately an order of magnitude smaller in magnitude. These findings suggest two key conclusions: for vortices pinned on the defects, the dominant force enhancement is along the x-direction, perpendicular to the defect lines; whereas for vortices pinned between the defects, the primary enhancement occurs along the y-direction, parallel to the defect orientation.
\par
Lastly, the line defects can be modeled using a Gaussian profile, enabling more precise estimations when the local Pearl length is extracted from susceptibility measurements.
\par
\section{Conclusions}
Overall, this work investigates the Lorentz force exerted by a scanning SQUID on vortices pinned in both homogeneous regions and regions adjacent to line defects in a thin-film superconductor. In homogeneous regions, the SQUID applies an isotropic dragging force on a vortex. For a vortex with a Pearl length $\Lambda = 420~\mu\text{m}$, the force magnitude is estimated to be on the order of a few piconewtons when the SQUID's effective field coil radius is 7.13~$\mu$m and positioned 4~$\mu$m above the film. An analytical expression is derived to calculate the Lorentz force on a vortex for arbitrary Pearl lengths and SQUID geometries, and this expression is shown to converge with an independent numerical method based on inverse Fourier transforms. The numerical method is then used to evaluate the dragging force on vortices pinned on and between two parallel line-shaped superfluid density defects. In both the x- and y-directions, the presence of such defects—which locally reduce the Pearl length—enhances the Lorentz force applied by the SQUID. Specifically, vortices pinned directly on the defects experience enhanced force primarily perpendicular to the defect orientation, while vortices pinned between the defects experience enhanced force parallel to the defect lines. These results enable more accurate estimation of Lorentz forces in the presence of structural inhomogeneities and contribute to a deeper understanding of vortex pinning and dynamics in superconductors with intrinsic defects, such as twin boundaries commonly found in high-temperature superconductors. The simulation framework can be readily extended to bulk superconductors and to defects with more general geometries.
\section{Acknowledgement}
This work was completed by R. A. Shi based on work during her graduate studies at Stanford University. She acknowledges Kathryn A. Moler for funding support and Yusuke Iguchi for assistance in verifying calculations.

\bibliographystyle{apsrev4-2}
\bibliography{apssamp}

\end{document}